\documentclass[lettersize, journal]{IEEEtran}
\usepackage{cite}
\ifCLASSINFOpdf
  \usepackage[pdftex]{graphicx}
  \DeclareGraphicsExtensions{.pdf,.jpeg,.png}
\else
\fi
\usepackage{amsmath}
\usepackage{algorithmic}
\usepackage{array}
\ifCLASSOPTIONcompsoc
\usepackage[caption=false,font=normalsize,labelfont=sf,textfont=sf]{subfig}
\else
  \usepackage[caption=false,font=footnotesize]{subfig}
\fi
\usepackage{amsfonts,amssymb,amsthm}
\usepackage{hyperref}
\usepackage{cleveref}
\hypersetup{
    colorlinks,
    linkcolor={red!50!black},
    citecolor={blue!50!black},
    urlcolor={blue!80!black}
}
\usepackage{comment}
\usepackage{tikz}
\usepackage{tikz-timing}

\newcommand{\signal}[1]{\texttt{#1}}

\begin{document}
\title{Experimental Verification of\\ Fast Voltage Droop Correction Circuits}

% \author{Shreyas Srinivas,~\IEEEmembership{Member,~IEEE,}
%         Ian W Jones,~\IEEEmembership{Fellow,~OSA,}
%         Carsten Schulze,
%         Milos Krstic,
%         Christoph Lenzen,~\IEEEmembership{Life~Fellow,~IEEE}% <-this % stops a space
%\author{\IEEEauthorblockN{Shreyas Srinivas, Ian W. Jones, Goran Panic, Christoph Lenzen}}
\author{
\IEEEauthorblockN{1\textsuperscript{st} Shreyas Srinivas}
\IEEEauthorblockA{\textit{CISPA Helmholtz Center for Information Security} Saarbr\"{u}cken, Germany
%\orcidlink{https://orcid.org/0000-0002-3993-1596}}\\
%shreyas.srinivas@cispa.de\\}
}\\
\and
\IEEEauthorblockN{2\textsuperscript{nd} Ian W. Jones}
\IEEEauthorblockA{\textit{}
Palo Alto, California, USA}\\
%ian.w.jones@ieee.org\\}
\and
\IEEEauthorblockN{3\textsuperscript{rd} Carsten Schulze}
\IEEEauthorblockA{\textit{IHP Leibniz-Institute for Innovative Microelectronics}
Frankfurt an der Oder, Germany}
%cschulze@ihp-microelectronics.com}\\
\and\\
\IEEEauthorblockN{4\textsuperscript{th} Milos Krstic}
\IEEEauthorblockA{\textit{IHP Leibniz-Institute for Innovative Microelectronics}
Frankfurt an der Oder, Germany}\\
%krstic@ihp-microelectronics.com}\\
\and
\IEEEauthorblockN{5\textsuperscript{th} Christoph Lenzen}
\IEEEauthorblockA{\textit{CISPA Helmholtz Center for Information Security}
Saarbr\"{u}cken, Germany}
\thanks{This work was supported in part by the European Research Council
(ERC) through the European Union’s Horizon 2020 Research and Innovation
Programme under Grants 716562 and 101123525.}}% <-this % stops a space
\maketitle
\begin{abstract}
    Due to the trend towards minimizing guard bands for energy saving purposes, voltage droops are a key limiting factor for the operational frequency of today's VLSI circuits. Adapting clock frequencies dynamically presents the challenge of metastability in the device that detects and stores the existence of voltage droops. We present an implementation of a fast all-digital circuit for adaptive response to droops using IHP's 130\,nm process. The description of the design is presented in an accompanying paper. We experimentally validate the functionality of the design on a test chip. 
\end{abstract}


% \fi
\IEEEpeerreviewmaketitle

\section{Introduction}
Fast voltage droops, i.e., drops in the supply voltage to a chip that play out within tens of clock cycles or less, are a challenging problem in the design of chips. Unchecked, the induced increased delay of circuit components leads to violations of timing constraints, which cause system failure. Sufficiently slow droops are countered by the control loop regulating the supply voltage. Faster droops, which occur within a few clock cycles, e.g.\ due to sudden load changes, kick in too fast for this control loop to respond in time. 

Traditionally, such faster droops have been addressed by sufficiently large guard bands in the supply voltage. Unfortunately, this approach is at odds with the goal of minimizing energy consumption, which necessitates to minimize the supply voltage and hence the size of the guardband. On the other hand, choosing the clock frequency conservatively to be safe even in the face of substantial droops is unsatisfactory due to the imposed limitation on clock speed.

A different approach is to slow down the clock frequency temporarily on the onset of a droop, to raise it again to the nominal value once the droop is over. Such an adaptive solution incurs a loss of performance, i.e., clock speed, only when necessary. In exchange, it introduces the challenge of recognizing and responding to a droop rapidly.

\paragraph*{Prior work on adaptive responses to voltage droops} We note that ours is not the only adaptive approach to voltage droops. In \cite{SVMDroopPredictor}, the authors use a fine-grained predictor of voltage droops based on known patterns of current draw caused by specific combinations of instructions in micro-architectures, and then altering the clock frequency sufficiently ahead of time. Such a solution is necessarily constrained to work on droops that can be predicted from specific common patterns. In \cite{dynamicvoltagecompensationHoltz, dynamicvoltage_second}, the proposed circuits detect voltage droops and compensate for current fluctuations via an additional higher voltage supply. As a slight variation, \cite{6422331} gate clocks when a voltage is detected. \cite{droopDNNs} focuses on the specific case of neural net architectures for deep neural net inference. In contrast our solution is architecture-agnostic. It focuses on dynamically adapting the clock signal frequency immediately after a droop is detected. Our response time is comparable to \cite{dynamicvoltage_second} i.e. within two clock cycles when accounting for the latency of droop detection.

A key obstacle to this is the threat of metastability, i.e., the possibility that a setup/hold time violation drives a latch or register into an unstable equilibrium state that is neither logic 0 nor 1. This naturally arises in digital approaches, as the onset of the droop need not be synchronized to the clock used to sample the voltage level, possibly resulting in an ambiguous measurement of whether or not the voltage level is considered too low. Using synchronizers to reduce the upset probability to acceptable levels incurs a delay of several clock cycles; thus, only relatively slow droops that leave several clock cycles from the point in time when they are detectable to causing critically low voltage levels can be covered. This prompted various solutions that rely on analog properties of the circuit elements. 

%While the details vary, these techniques have in common that they need to be carefully tuned to the technology node, rendering them high-effort tailored solutions that do not readily transfer to the next chip generation. We refer to~\cite{fugger2021fast} for a more detailed discussion.

\paragraph*{Metastability-containment}
F\"ugger et al.~\cite{fugger2021fast} proposed an all-digital circuit based on the paradigm of metastability-containment. Roughly speaking, they propose to ``mask'' internal metastability of registers using high- or low-threshold inverters until it resolves, reducing the effect of metastability to a (possibly arbitrarily) late output transition of the register. Combining this with careful leverage of masking properties of logic gates, essentially they create a synchronizer chain, but manage to use the values propagating through it right away to decide whether to increase the clock period or not.

A limitation of their work is the use of high- or low-threshold inverters. Especially under conditions of varying supply voltage, this is likely to require a much larger guardband. This defeats the purpose of an adaptive response to droops. Srinivas et al.~\cite{srinivasDesign24} address this issue by presenting a variant of the design that uses masking latches based on differential sensing. Moreover, they introduced improvements that remove dependency of the circuit on accurate control over delays and fully specify a design for IHP's 130\,nm technology.

\subsection*{Our Contribution}

We implemented the design proposed in~\cite{srinivasDesign24} and verified its functionality experimentally. This article presents the experimental setup and choices alongside our findings. We also detail the rationale behind the experiments and their limitations. In particular, we discuss why both qualitatively and quantitatively, it is impractical to experimentally obtain useful data of metastability within the circuit. Nevertheless, we accomplish the next best thing possible: we a range of voltages to the droop detection input that might reflect the output of a metastable latch from whose output Q, this input is produced for short and long durations. For the short duration droops we present the analog and digital waveforms of our clocks. For the long duration signals we collect clock frequencies which demonstrably indicate that our chips output correctly slowed clocks when a droop occurs.

%\subsection{Previous Work}
%In their work on fast clock adaptation circuits for handling voltage droops \cite{fugger2021fast}, the authors presented a flip flop based design for an augmented synchroniser chain that handled an incoming voltage signal with a potential droop. Their design was built on the assumption of availability of flip flops which exhibit metastability containment. Specifically, they assumed a flip flop whose voltage thresholds were altered to output a 0 or a 1 depending on the designer's configuration, during the occurrence of internal metastability in the contained latches. Further their delay elements are rather brittle to variations in delay caused by temperature and process changes. They justify this brittleness by introducing phase-locked-loops (hereon PLLs) to stabilise all their delay lines. Despite this, the authors of \cite{srinivasDesign24} uncovered a few discrepancies in their design. In the work \cite{srinivasDesign24}, fixed these discrepancies and presented a design of metastability containing \emph{latches}. This resulted in a circuit whose synchroniser chain was halved in length, at the cost of additional complexity in the proofs ensuring correctness of their design. This work presents the results of the implementation of the design of \cite{srinivasDesign24} on a test chip using the IHP 130 nm CMOS technology.

\begin{figure*}[t!]
    \centering
    \includegraphics[width=0.6\textwidth]{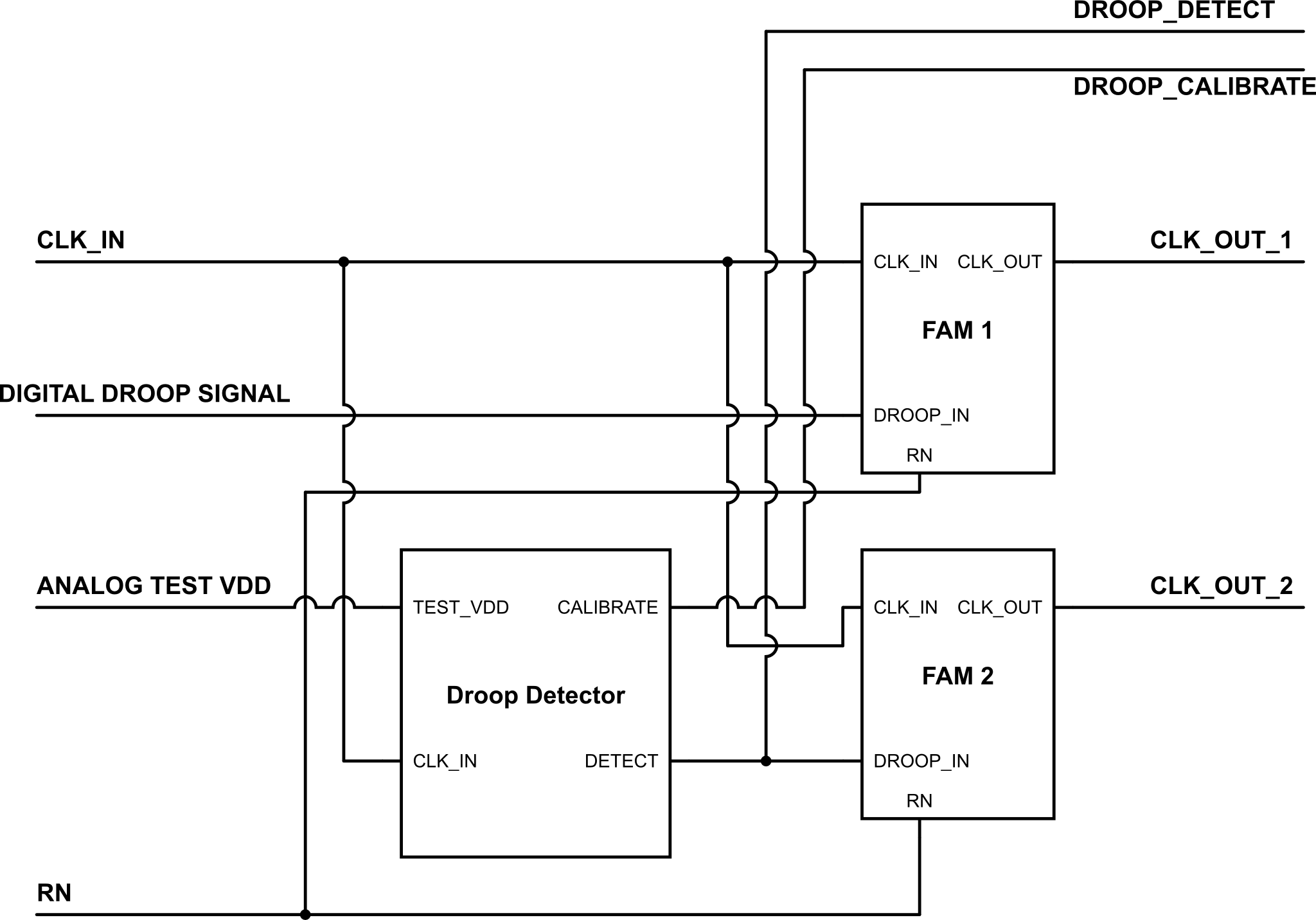}
    \caption{The top level schematics of the test chip. Note that digital and analog refer to the I/O port type used, i.e., ``digital'' does not imply that the chip must receive a ``clean'' logic signal via the given port. RN is the reset signal}
    \label{fig:top-level-schematic}
\end{figure*}
\paragraph*{Organization of the Paper} \Cref{sec:setup} discusses the experimental setup, describing the test chip and characterizing the testing capabilities we had at our disposal. A detailed description of the tested design along with proofs of correctness adapted from \cite{fugger2021fast} can be found in \cite{srinivasDesign24}. In \Cref{sec:experiments}, we discuss the goals of our experimentation and the derived experiments, including the expected results. In particular, we explain why studying metastability within the circuit faces both fundamental challenges and such that are imposed by the experimental setup. 
Last, but not least, in \Cref{sec:results} we summarize our findings in the experiments, compare them against expectations. The figures presented in this section represent a sampling of our results. The supplementary material accompanying this paper contains the data for clock frequency measurements under long droops. Finally we conclude in \Cref{sec:conclusion}.

\begin{table}[t!]
  \centering
  \caption{The pins of our test chip.
  \label{table:toplevel_pinlist}}
  \begin{tabular}{|m{2.8cm}|m{0.8cm}|m{0.8cm}|m{2.63cm}|}
    \hline
    \textbf{Pin} & \textbf{In/Out} & \textbf{Type} & \textbf{Purpose} \\ \hline
    RN  & in & digital & global active-low reset\\ \hline
    CLK\_IN & in & digital & 400 MHz clock\\ \hline
    DIGITAL DROOP SIGNAL & in & analog &  digital droop detection input for FAM\_1 \\ \hline
    ANALOG TEST VDD & in & analog & artificial droop VDD for droop detector\\ \hline
    DROOP\_CALIBRATE & out & digital & calibration output of droop\_detector\\ \hline
    DROOP\_DETECT & out & digital & droop detected output for the Droop detector\\ \hline
    CLK\_OUT\_1 & out & digital & clock output of FAM\_1 \\ \hline
    CLK\_OUT\_2 & out & digital & clock output of FAM\_2\\ \hline
  \end{tabular}
\end{table}

\section{Experimental Setup}\label{sec:setup}
\subsection{The top level structure of the test chip}
At a high level, our design consists of two pieces as shown in \Cref{fig:top-level-schematic}:
\begin{itemize}
    \item The Frequency adaptation module (FAM): This module responds to a droop detection signal and acts upon it to produce slow or fast clock pulses.
    \item The Droop Detector: This module is our droop detector as designed in \cite{srinivasDesign24}. It receives an analog voltage signal as input, which acts as our test signal. It is in this signal that we artificially induce droops for experimentation. We note that this droop detector is included on chip purely for purposes of completeness and it remains to be seen whether this design can be improved upon.
\end{itemize}
We emphasize that the design assumes that the relatively small circuitry implementing these components is \emph{not} subject to droops in its supply voltage. This is justified by its small and stable power usage and the possibility to supply it with power via separate pins, or in a large system even a fully separated power supply, to shield it from the effects of strong load changes in the application circuitry it supplies the clock signal to.

The test chip is designed to experimentally verify the response of our design to the occurrence of voltage droops. Specifically, we wish to observe the output clock signals from our chip given an input clock signal and an artificially controlled supply voltage reference presented to the detector subcircuit. The result of our tests are superimposed samples from multiple runs of a tester for each given test bench. We would like to test the FAM module independently of the droop detector and vice versa, as well as the complete design as a whole. Thus, we put two copies of the FAM module in the test chip.
\begin{itemize}
    \item One FAM module is completely disconnected from the droop detector and supplied with external signals only. It receives an input clock, a droop detection signal, and a reset signal. Its sole output is the clock output, which is connected to an output port.
    \item A second FAM module is accompanied by the droop detector. Its droop detection input is produced by the droop detector, which also produces a calibration output. 
\end{itemize}
\Cref{table:toplevel_pinlist} lists the I/O pins of the chip.

\begin{figure}[t!]
    \centering
    \includegraphics[width=\columnwidth]{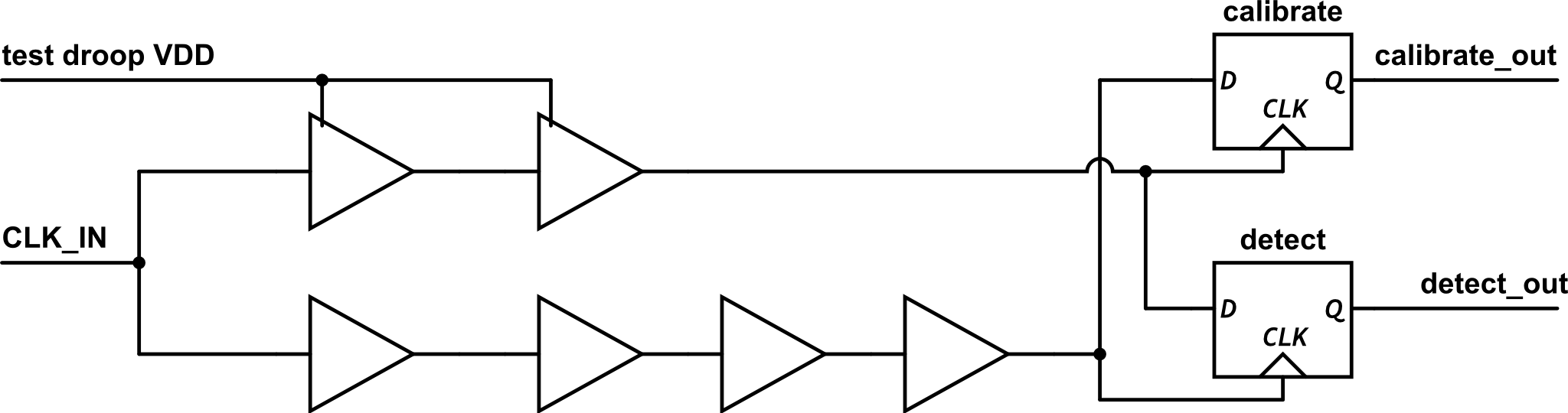}
    \caption{Schematic of our droop detector.}
    \label{fig:droop-detector-schematic}
\end{figure}

\begin{figure*}[ht]
    \centering
    \includegraphics[width=0.7\textwidth]{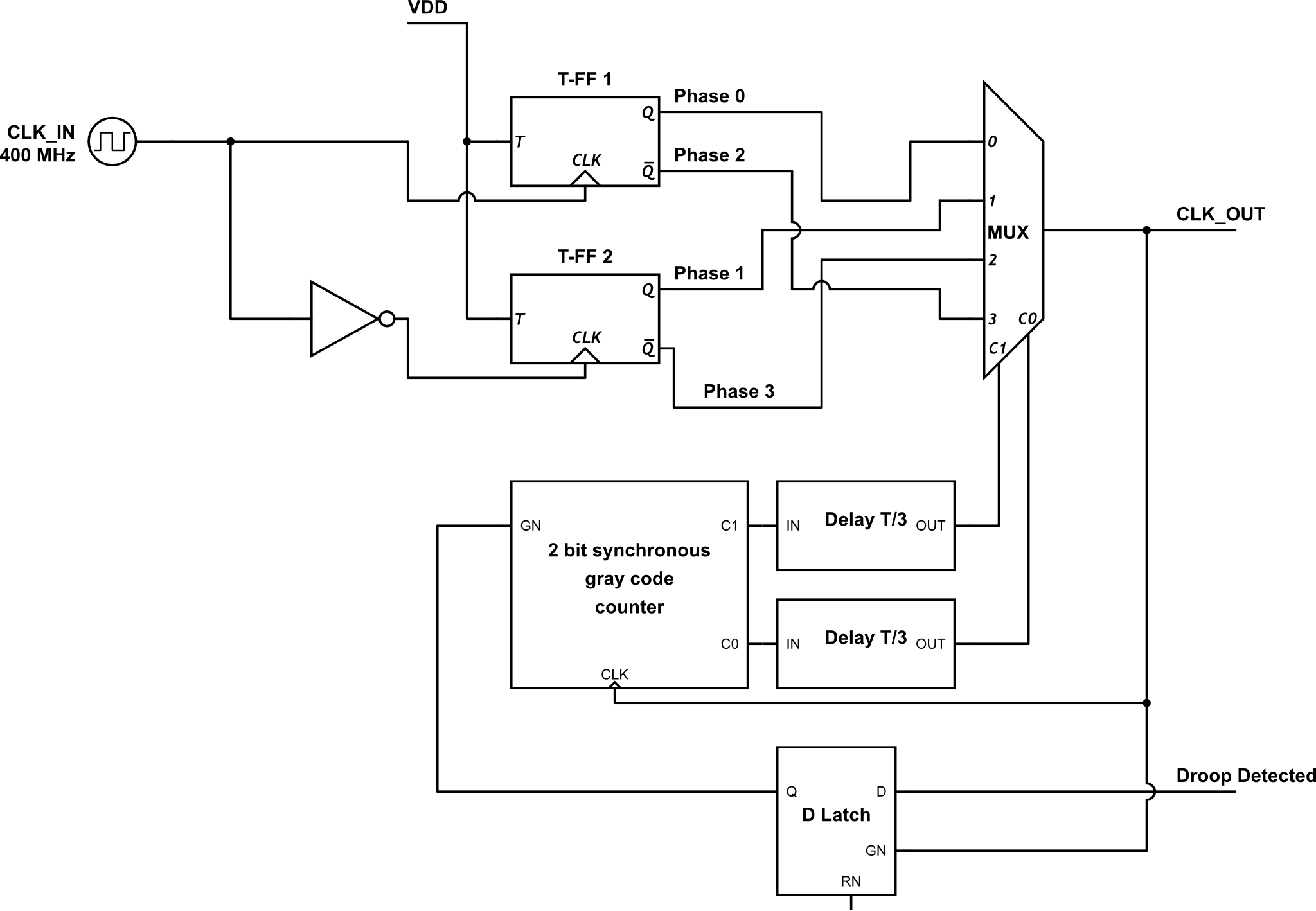}
    \caption{The phase accumulator of our circuit. The figure originally appears in the paper that describes our design\cite{srinivasDesign24}}
    \label{fig:phase_acc_schematics}
\end{figure*}

\begin{figure*}[ht]
    \centering
    \includegraphics[width=0.7\textwidth]{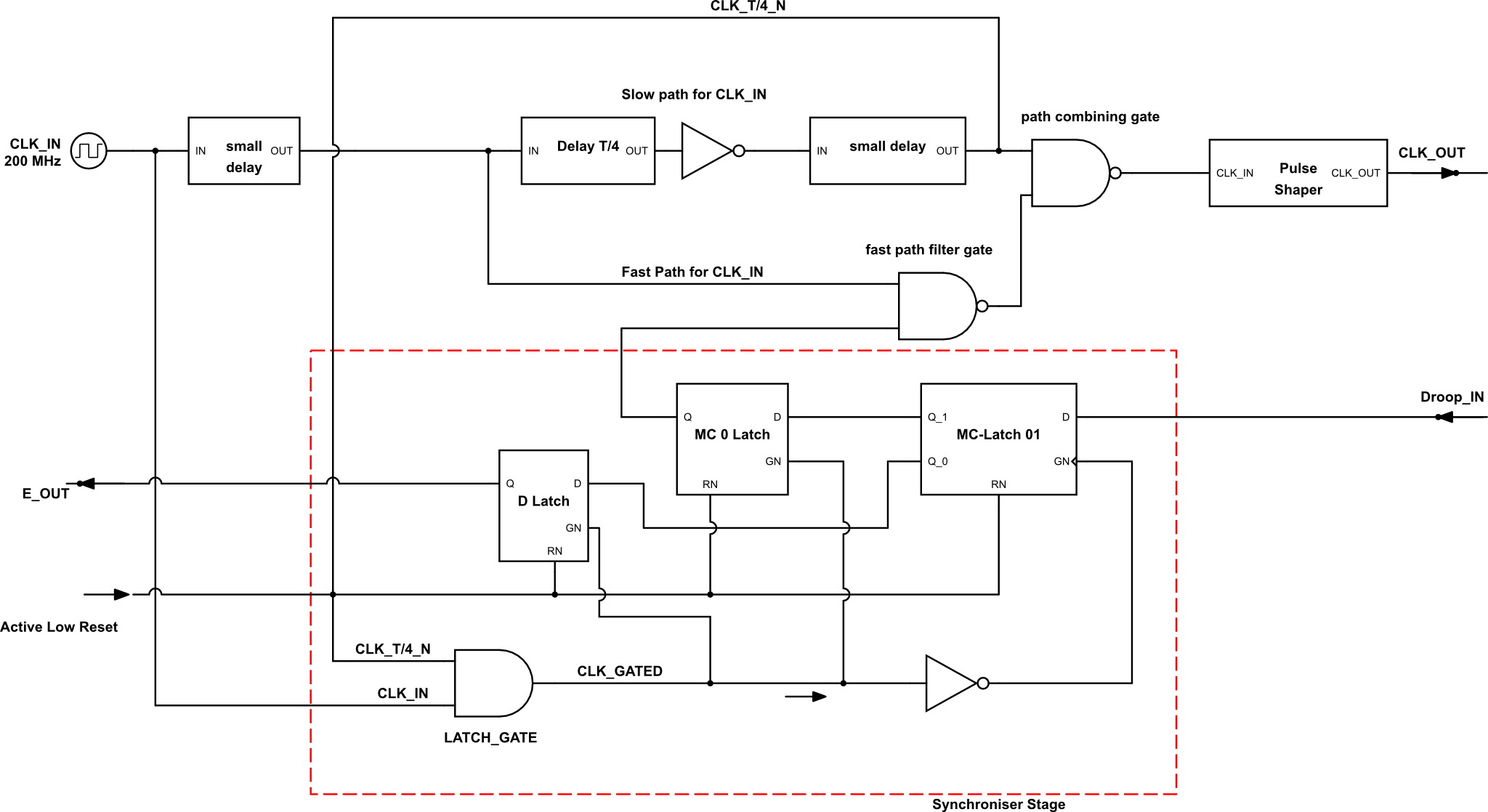}
    \caption{The delay element in the basic unit of the adaptive synchronizer chain of our circuit and is borrowed from our paper that goes deeper into the design and correctness issues of this circuit~\cite{srinivasDesign24}. The highlighted part in red depicts the latches that provide the functionality of a single synchroniser stage, while passing on the droop signal to a circuit that will pre-emptively add a phase offset to one clock pulse if a droop is unambiguously detected.}
    \label{fig:delay_elem_schematics}
\end{figure*}

\subsection{The Basic Parameters}
Our chip operates on a supply voltage of $1.4V$ and ground voltage of $0V$. For our experiments  particular with the droop detector and \signal{FAM\_2}, we supply a test droop supply voltage through an analog port. We vary this voltage between $0.9V$ and $1.3V$. This supply voltage supplies the test line of our droop detector through its \signal{test droop VDD} input port (see \Cref{fig:droop-detector-schematic}).

We further supply one input clock for both our FAM modules through a digital input port. Keeping in mind the limitations of the frequency of input and output clocks, we originally designed our circuit to operate with \signal{CLK\_IN} with a frequency of $400$\,MHz. However, this proved to be at the very limit of the capabilities of our input ports, which resulted in very noisy output signals. Further our droop detector required this higher voltage to provide clean outputs. Accordingly, we instead performed our experiments by supplying \signal{CLK\_IN} with a square clock of frequency of $333$\,MHz. 

\subsection{Technology specific design adaptations}
As explained in \cite{srinivasDesign24}, we adapt our design to work without Phase Locked Loops (PLLs), by using T Flip Flops to generate various phase offsets of our input clock. This results in a halving of the clock frequency at the output of FAM 2. Further, in order to fall safely within the limits of our clock output ports, we halve the output frequency further. Thus, the ports \signal{CLK\_OUT\_1} and \signal{CLK\_OUT\_2} produce clocks whose output frequencies range between $80$ and $100$\,MHz, depending on our artificially provided droop inputs. As we shall explain below, operating close to the frequency limits of our input and output ports resulted in a very noisy output. Fortunately, our design is resilient to a 16.7\% lower frequency clock input. Thus, in the end, our experiments used input clocks of frequency 333 MHz and output clocks of frequency about $66$ to $83$\,MHz. Nevertheless, we emphasize that these modifications are an artifact of designing a test chip geared for experimental observation. The only crucial frequency halving to achieve the desired functionality is the one performed by our phase accumulator, which generates the four phase offsets of our clock without using PLLs. Moreover, due to the small depth of its logic, in typical designs the frequency adaption circuit will be able to run at a substantially higher frequency than the logic it provides the clock signal for.

The delay elements, which are the key components of our FAMs, are sensitive to their input clock frequency. This is because we use delay chains whose delays are fixed fractions of the nominal time period. Thus changing the frequency as above could potentially affect the correctness of our design. Anticipating the effects of PVT variations, as explained in  \cite{srinivasDesign24}, we took a conservative approach to our delays in the delay elements of our FAMs. This enabled us to handle the modified input clock frequencies. The trade off was a slightly unbalanced duty cycle in our our output clock pulses when running at the reduced frequency. 

\subsection{Testing Tools and Methodology}
In this section, we briefly explain the details of our test setup, and specifically how we collect the traces of the output clock in our experiments.
\begin{itemize}
    \item For the testing we use an Adventest v93000 digital tester. It enables us to automatically sweep through the voltage of pins separately and to measure frequency of the output clocks over some cycles.
    \item The test method is written in C++ with the help of the tester-API. It measures timing events, when a trigger voltage (1.1V) is crossed. With this, the frequency of the signal, in our case CLK\_OUT\_1 and CLK\_OUT\_2, can be calculated. The arming of the signals starts after changing the voltage of droop\_VDD and droop\_deteced.
    Each pin has three voltage levels: low (default 0V), high(default 1.4V) and Z (default 0.574V), which are changed in a loop, to observe the difference in frequency at the clock-output pads.
    \item At the beginning of the testflow, the basic function of the die is tested, via Continuity and short tests. After that functionality of the basic clock feature is checked.
    At the end the different loops, realized with so-called "Shmoo"-tests, are performed.    
\end{itemize}

The output of our tests is a series of waveforms produced on the software interface of our tester. Additionally we also measure the clock frequency of the output clock through scoped sampling.

%Yes, I try to explain: 
The digital tester does not operate like a standard oscilloscope when drawing the timing diagrams. Since the tester is operating at a defined frequency and voltage at any given time; it does not over-sample the output of the chips. Instead, it measures the signal with a shifted timing and different voltages multiple times. Therefore, in order to create the images, the test-pattern must be repeated multiple times, and only deterministic timing behavior over these repetitions is observed as a stable signal.

The output waveforms are recorded in one of two forms: analog and digital. These waveforms are reconstructed by the tester by running each test multiple times and strobing the output waveform from sampling the output in each run.

\section{Goals of our Experimental Effort}\label{sec:experiments}
In this section we describe the experiments we conduct on our test chip and the results we obtained. In each experiment, we look as the waveforms of the output clocks produced by the testing machine, \signal{CLK\_OUT\_1} and \signal{CLK\_OUT\_2}, as we vary \texttt{droop\_VDD} and \texttt{droop\_detected}. These trace outputs are obtained by running the test bench multiple times and sampling the output signals at a high frequency.

\subsection{What we test for}
At a basic level, to confirm the correctness of our circuit, we seek to test three aspects of our test chip's functionality:
\begin{itemize}
    \item \textbf{Behavior in standard operation:} We verify that under normal operating conditions, when the FAM modules sample stable values from their \signal{DROOP\_IN} inputs, the output clocks behave as expected: they slow down when the signal indicates a droop and run at regular speed when it does not.
    \item \textbf{Behavior under non-logic input:} When the droop detection input is at an ambiguous voltage level, how does our circuit respond? The expected outcome is one of the following: (i) the input is interpreted as ``no droop detected,'' i.e., the default output clock speed is generated, (ii) the input is interpreted as ``droop detected,'' i.e., the reduced output clock speed is generated, or (iii) some rising clock transitions are delayed, possibly less than the full amount for the increased period, while others are not. Nonetheless, minimum and maximum period length are observed and all output clock transitions are ``clean,'' i.e., there are no glitches or slow signal transitions. Observing option (iii) would imply that metastability in the latches of the delay elements has been induced. However, we do not directly observe internal voltage levels, so we also cannot directly determine whether metastability occurred.
    \item \textbf{Functionality of our droop detector design:} Lastly, we would like to confirm that our droop detection circuits work as expected. Given that we use long delay lines, a droop detected by them is only transferred to the frequency adaptation module (FAM II) after a clock cycle. We emphasize that we do not believe this to be the best achievable speed for voltage droop detection. Rather, the goal is to demonstrate \emph{some} complete implementation of the frequency adaption circuit without investing significant effort into a technology-dependent analog (or quasi-analog) component.
\end{itemize}

\begin{figure}[t!]
    \centering
    \input{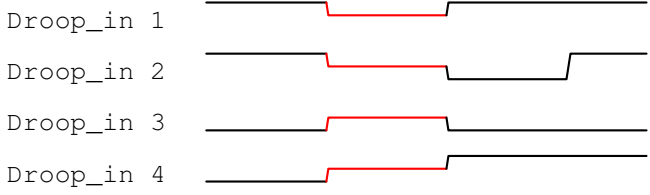}
    \caption{Different forms of the droop input signal we use as a test bench. We vary the voltage for the red segments of the signals. The first test bench represents the case where no droop is detected after some potential metastability, and the second test bench denotes the case where the droop detection input goes metastable but ultimately stabilizes to indicate no droop. In the experiments, these symbols were combined into one long signal.}
    \label{fig:testbench-droop-signal}
\end{figure}

\begin{figure*}[t!]
    \centering
    \includegraphics[width=0.7\linewidth]{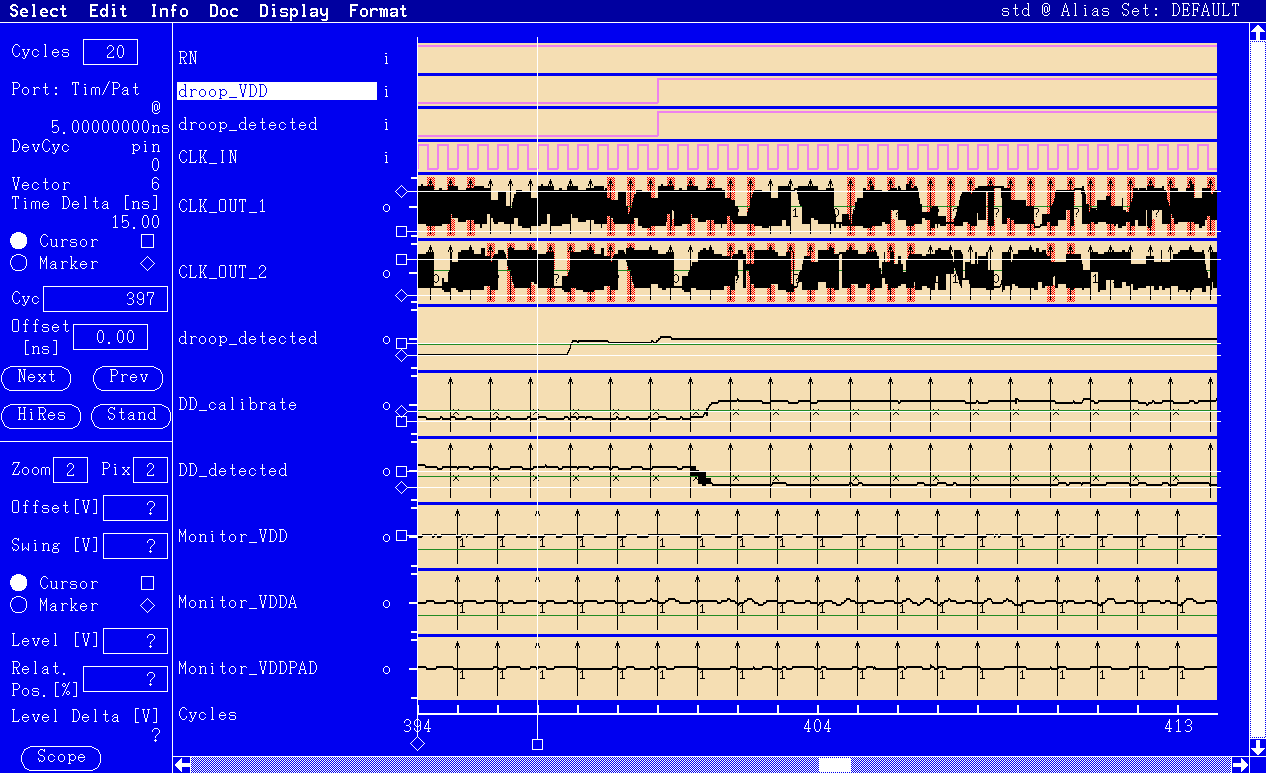}
    \caption{An example result from our tester tool. The signal \signal{CLK\_IN} is the input clock signal shown as a digital input. The respective output clocks of 
    \signal{FAM\_1} and \signal{FAM\_2} are \signal{CLK\_OUT\_1} and \signal{CLK\_OUT\_2}. The time range marker denoted by white lines marked with a diamond and a box at the bottom indicate a range of time, which in this case spans 6 clock cycles of the \signal{CLK\_IN}. As the value of \signal{Time Delta} field on the left hand side panel indicates, this is a 15 nanosecond interval. This translates to 2.5ns per clock cycle, which indicates that the input clock \signal{CLK\_IN} has a clock frequency of 400 MHz. The extremely noisy outputs of \signal{CLK\_OUT\_1} and \signal{CLK\_OUT\_2} indicate that we are at the limit of the capacity of our IO ports.} 
    \label{fig:noisy_out_bad_frequency}
\end{figure*}

\subsection{Our test benches}
In order to ensure that we can separately analyze the behavior of the FAM module, as well as its combined behavior with the droop detector, in principle we could experiment separately on two inputs. In practice, the two test benches are combined into one, since the two copies of the frequency adaptation module can be tested concurrently, while also getting outputs from the droop detector. Thus, the reader will observe that the droop inputs as well as both our clock inputs change frequencies very closely aligned in time with each other. Secondly, we combine testing under stable and ambiguous inputs in \signal{FAM I} as follows: We have a combined test bench where \signal{droop\_detected}, which is active low, steps down from the inactive logic level, \signal{logic 1}, to an intermediate voltage which we call $V_{\mathrm{int}}$ and then to the active \signal{logic 0} level. While returning to the inactive state, the signal jumps back to \signal{logic 1}. The shape of the test bench input signal for the droop inputs is shown in \Cref{fig:testbench-droop-signal}. We accomplish this in our digital tester, which only accepts standard logic values as inputs, by giving it a \signal{logic Z} input which we manually configure to a voltage equal to $V_{\mathrm{int}}$.

In our technology, $V_{\mathrm{DD}} = 1.3$\,V and $\mathrm{GND} = 0.0$\,V. We vary $V_{\mathrm{int}}$ between $0.5$\,V and $0.6$\,V, with the goal of identifying a voltage that from a digital point of view, marks the transition between reading as a \signal{logic 1} and or \signal{logic 0}. Around that voltage, which might vary depending on PVT variations, we anticipate a noisy output from our strobing sampler indicating that internal metastability of sampled \signal{DROOP\_IN} values was induced. We remark that we have no way to confirm whether this is truly the case in our test setup, however. Lastly, we will re-use some of the images to explore the results of multiple experiments. Since our experiments do not collect statistical information, but serve to demonstrate the qualitative behavior of our chip, this does not result in any statistical issues.

\subsection{Test Outputs}
 We collect two forms of output from our experiments. 
\begin{itemize}
    \item Digitally reconstructed waveforms of the output clocks from our tester.
    \item Clock frequencies of the output clocks at specific segments when a droop is or is not asserted by the corresponding input signal.
\end{itemize}
% We conducted our tests on a digital tester which uses strobing

%\TODO{Is \emph{strobing} the right word? (Remark Carsten: In principle yes, it is the right word. But sounds a bit unexpected here. Therefore I tried to explain: 

In order to draw the timing diagram the tester strobes the signal at different time points for different expected VOL(Voltage output low) and VOH (Voltage output high), the respective voltages of logic 0 and logic 1, to be able to reconstruct the signal. So when a signal does not have a repeatable/stable voltage in the pattern (for example, because of missing reset), it will create black bars. Either a different method must be used or the pattern must be changed in a way to have predictable outputs. If there are clear lines, it shows that the test is repeatable and the voltage is stable. Thus each test bench is run multiple times and the output signal is reconstructed as the overlapped version of the output signals sampled from each run. Further our tester allows us to measure the frequency of our output clocks for specific time segments. The waveforms allow us to confirm that there were no glitches in the output clocks, including when transitioning between the slow and fast output frequencies.

\subsection{Running our circuits at the limits of our ports}
Recall that we deviate from our original experimental design in that our chips are tested with clocks of input frequency 333\,MHz, while the original design called for testing at 400\,MHz. This higher frequency, along with the corresponding output frequency of up to 100\,MHz, is at the limit of the capabilities of our input and output ports, respectively. The result is a very noisy output on both clocks. This behavior is demonstrated in \Cref{fig:noisy_out_bad_frequency}. Fortunately, our design is resilient to reducing the input clock frequency to 333\,MHz, due to an entirely unrelated design decision to be conservative in the length of our delay lines to accommodate PVT delay variations. Hence, in all subsequent experiments, the frequency of \signal{CLK\_IN} is set to 333\,MHz, which corresponds to a time period of 3.03\,ns. Internally the clocks supplied to the \signal{FAM}s are 165 Mhz. Their frequencies are further halved to be within the allowed limits of the output port. Thus in our experiments, we expect to measure time periods of 80 MHz for the regular clock and 66 MHz for the slowed down clock.

\subsection{Why is deep metastability of the delay chain not explored?}
Standard upper bounds on the probability to resolve metastability of a latch in $t$ time are of the form $pe^{-t/\tau}$, where $p$ is the probability for a setup/hold time violation caused by a data signal transition close to latching and $t$ is the resolution time constant of the latch~\cite[Chapter 5]{kinniment2008synchronization}. Given that we are not in a position to publish the relevant numbers, let us perform a ballpark estimate of the time required to observe an upset.
\begin{itemize}
    \item As we would be actively trying to induce metastability using ``bad'' inputs, we cannot readily use standard MTBF formulas from the literature, which assume that data and clock transitions are statistically independent. On the other hand, we do not have sufficiently fine-grained control over experimental conditions to reliably drive the circuit into arbitrarily deep metastability. Rather, we can assume that we can get very close to the necessary conditions for extended metastability, but around that point the input distribution is close to uniform due to temperature and voltage fluctuations. To reflect this, let us pretend that $p=1$, but the exponential decay of metastability due to the $e^{-t/\tau}$ term is retained. We might be a few orders of magnitude off, but get a general idea of what to expect.
    
    \item In~\cite{srinivasDesign24}, simulations of the designed masking latches put the time constant $\tau$ of latches based on PTM 130nm as almost identical to that of a standard synchronizer latch, so the same is to be expected in IHP's technology.
    \item Our goal is not to analyze a specialized, highly technology-dependent solution. Hence the used latches exhibit the synchronizer performance of standard digital 130\,nm IHP technology. Even in older, less mature processes, $\tau$ was ranging under $100$ps as measured by \cite{tau_decrease_fails_below_100}. A modern process is likely to perform at least as good, so let us use $300$\,ps as a highly conservative upper bound.
    \item At the low rates of upsets to be expected, we must avoid noise on the output. Hence, we must run the circuit at $80$\,MHz output frequency or slower, i.e., the clock period is at least $T\geq 12.5$\,ns.
    \item Due to a conservative correction of a timing issue for the latch-based design revealed in~\cite{srinivasDesign24}, a sampled value is stored for more than three quarters of a clock cycle before it is used to decide whether to delay a clock pulse by the last element in the chain, i.e., $t>3T/4\approx 9$\,ns.
    \item Thus, $e^{-t/\tau}<e^{-30}\approx 10^{-13}$. Even when making an attempt every clock cycle, i.e., every $12.5$\,ns, it would in expectation take more than a day to observe a single event.
\end{itemize}
From this rough estimation it should become clear that it is impractical in terms of (expensive) measurement time, challenges regarding distinguishing ``true'' events from noise, and complexity of experimental setup to observe the output behavior of the circuit under deep metastability.

Alternatively, one might consider attempting to directly monitor the state of the internal storage loop of a delay element latch. Accessing this information requires either highly advanced probing equipment that is not at our disposal or ``copying'' the internal signals for the purpose of monitoring. In addition to necessitating a redesign, the latter would add load and thereby change the characteristics of the circuit elements whose behavior we would like to study. 

All in all, far-reaching changes to the experimental setup, e.g.\ massive parallelization or adapting the circuit to operate at much higher clock speeds, would be necessary to obtain useful data on the circuit behavior under extended metastability. This was beyond the scope of the project. Instead, this case was extensively studied in simulation before fabrication, prompting the timely discovery of the aforementioned timing issue discussed in~\cite{srinivasDesign24}.
\begin{figure*}[ht]
    \centering
    \includegraphics[clip, trim=0.3cm 10cm 0.5cm 10.9cm, width=1.0\linewidth]{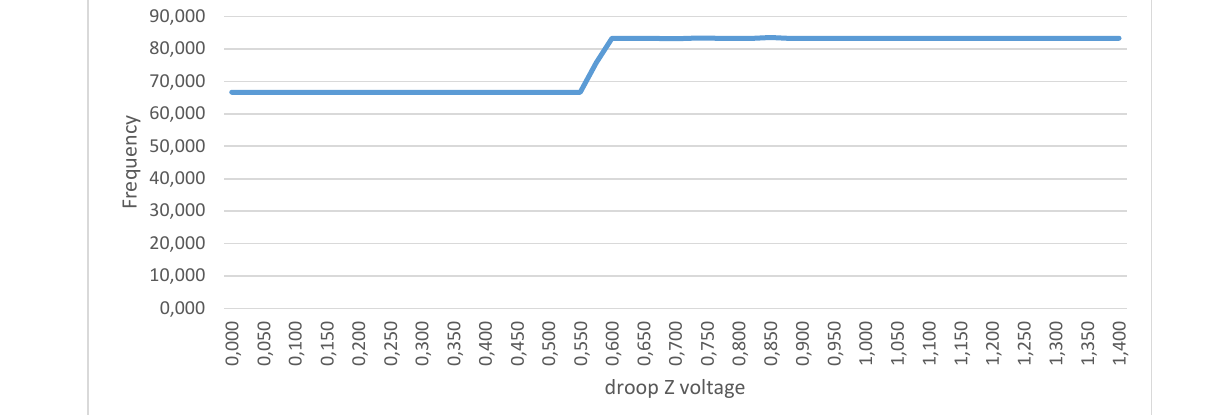}
    \caption{Clock frequency measured for a long droop plotted as a function of the droop voltage at the \signal{droop\_detected} input. This is accomplished by varying the voltage value of the Z voltage signal. The droop detection signal is asserted as the Z value for an extended time and clock frequency measured as the average of several runs of the test bench. In the graph, the value of the voltage for the high impedance logical signal is varied between 0.0 and 1.4 V in step sizes of 0.5 V. A transition in the clock frequency is observed when this voltage crossed from 0.550 to 0.650.}
    \label{fig:droop-detector-rise}
\end{figure*}

\begin{figure*}[ht]
    \centering
    \includegraphics[clip, trim=0.3cm 10cm 0.5cm 10.9cm, width=1.0\linewidth]{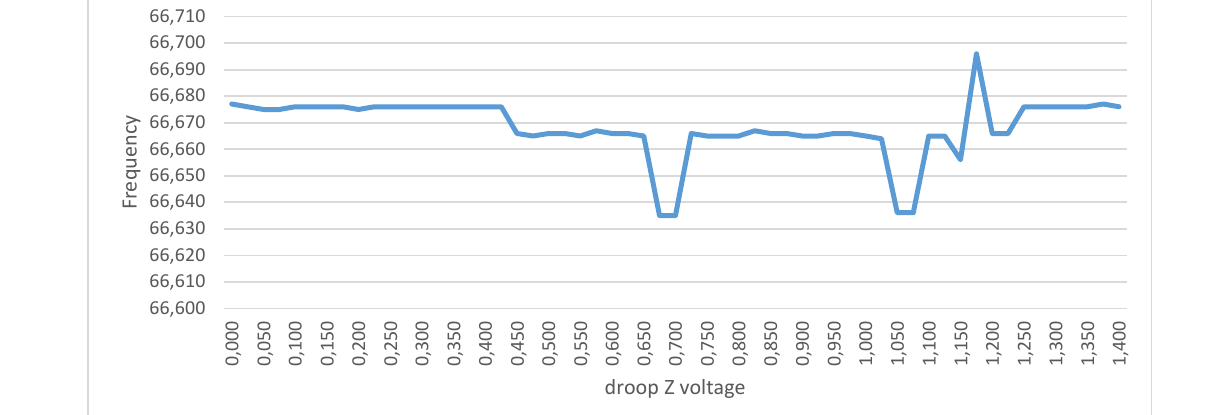}
    \caption{Clock frequency measured for a long droop plotted as a function of the droop voltage at the \signal{droop\_detected} input. Here the clock frequency measurement is rather stable since the frequency is measured when the droop is asserted. Note that the clock frequency variation on the y-axis is between 66.6 ns and 66.71 ns. The clock frequency variations observed are in fact nothing more than experimental variations.}
    \label{fig:droop-detector-clock-freq-1}
\end{figure*}

\begin{figure*}[ht]
    \centering
    \includegraphics[clip, trim=0.3cm 10cm 0.5cm 10.9cm, width=1.0\linewidth]{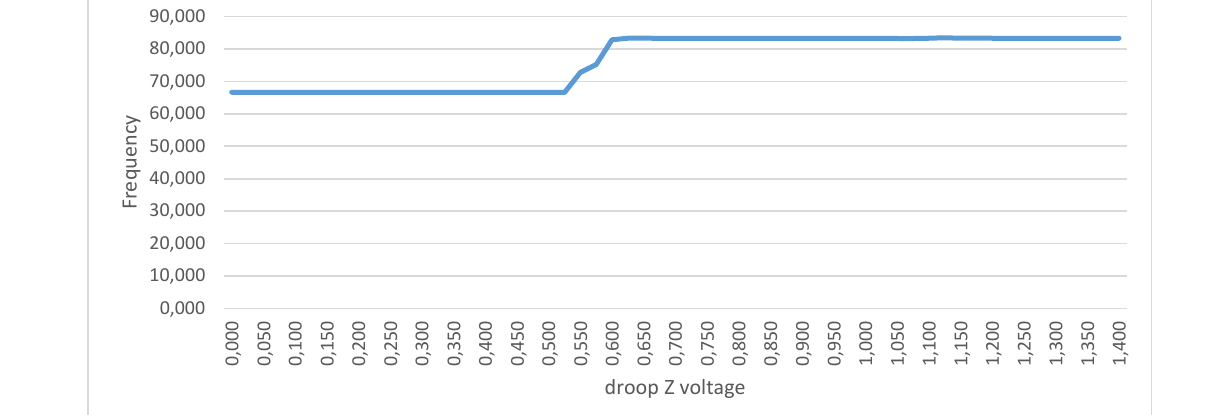}
    \caption{Clock frequency measured for a long droop plotted as a function of the droop voltage at the \signal{droop\_detected} input. This is accomplished by varying the voltage value associated by the digital tester with the high impedance logic value Z}
    \label{fig:droop-detector-clock-freq-2}
\end{figure*}

\begin{figure*}[ht]
    \centering
    \includegraphics[clip, trim=0.3cm 10cm 0.5cm 10.9cm, width=1.0\linewidth]{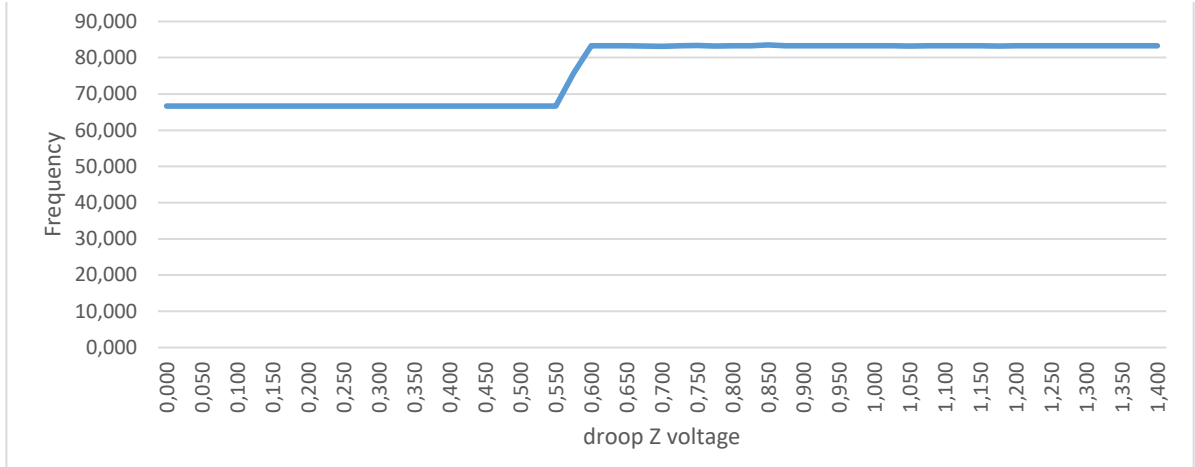}
    \caption{Clock frequency measured for a long droop plotted as a function of the droop voltage at the \signal{droop\_detected} input. This is accomplished by varying the voltage value of the Z voltage signal}
    \label{fig:droop-detector-clock-freq-3}
\end{figure*}

\section{Results}\label{sec:results}
In this section, we present the execution and outcomes of our experiments. For some tests we present the digital and analog traces from our tester. Further we show some sample plots for the frequency measurments of our output clocks under long droops. The full data set is available as part of the supplementary material.

\subsection{Testing the Droop Detector}
In our first tests, we check the functionality of our droop detectors. 

Let us recall the operational idea behind the droop detector circuit already presented in the companion article\cite{srinivasDesign24}. As shown in \Cref{fig:droop-detector-schematic}, the droop detector consists of two delay lines of buffers, which are connected to the data and clock inputs of two flip flops named \signal{calibrate} and \signal{detect}. Their respective outputs correspond to the signals \signal{DD\_calibrate} and \signal{DD\_detected} in \Cref{table:toplevel_pinlist}. The buffers on one line get the normal supply voltage of 1.4V from the supply lines of the chip. The buffers on the other line receive their supply voltage from an analog input signal we control called \signal{test\_droop\_VDD}. For the purposes of this discussion, we call the former the \emph{control line} and the latter the \emph{experimental line}. The experimental line is two buffers shorter than the control line. As long as \signal{test\_droop\_VDD} and \signal{VDD} are identical, the experimental line delays signals two buffer delays less than the control line. Thus when an identical clock signal is passed through both delay lines, at the \signal{calibrate} flip flop, each rising edge arrives two buffer delays earlier at the \signal{CLK} input than at the \signal{D} input. The reverse happens at the detect input. Thus when there is no droop in \signal{test\_droop\_VDD}, the output of the \signal{calibrate} flip flop is \signal{logic 1} and the output of the \signal{detect} flip flop is \signal{logic 0}. This denotes the absence of the droop. When a sufficiently large voltage droop is introduced in \signal{test\_droop\_VDD}, the experimental line's delay extends beyond the delay of the control line. Now the order of arrival of rising edges at the clock and data inputs of both flip flops is reversed. Thus, when a droop is introduced, the \signal{calibrate} flip flop outputs \signal{logic 0} and the \signal{detect} flip flop outputs \signal{logic 1}. 

\Cref{fig:droop-detector-clock-freq-1} shows representative test results when the standard supply voltage is 1.4V. The test input to the droop detector, \signal{test\_droop\_VDD} is derived from the input port \signal{droop\_VDD}, which is an analog port. Its timing diagram is represented digitally in \Cref{fig:droop-detector-clock-freq-1}. The cursor is positioned in the middle of the timing diagram as observed by the box marker at the bottom. At the 200th clock cycle of \signal{CLK\_IN}, (which is 1200ns into the test), \signal{droop\_VDD} drops to 1.0V. At approximately the 201st input clock cycle, the signal \signal{DD\_calibrate} falls as expected. Recall that the frequency of the internal clock is half the input clock frequency. From this point on, we call it the \emph{operational} clock frequency of the circuit. Thus, from an operational standpoint, the transition occurs within half a clock cycle. The \signal{DD\_detected} rises at the same time, but then unexpectedly falls again for half an operational clock cycle until finally settling to the correct stable output. Because this is observed consistently across our tests, metastability is unlikely to be a cause. We suspect a timing flaw in the droop detector's setup, which manifests moreso when the normal supply voltage is set at 1.2V, but we lack a fully satisfactory explanation for this behavior. Nevertheless, our droop detector's outputs settle into their correct values within two operational clock cycles, and in principle one could make use of the negated \signal{DD\_calibrate} signal instead of the \signal{DD\_detected} signal.
\begin{figure*}[t!]
    \centering
    \includegraphics[width=0.7\linewidth]{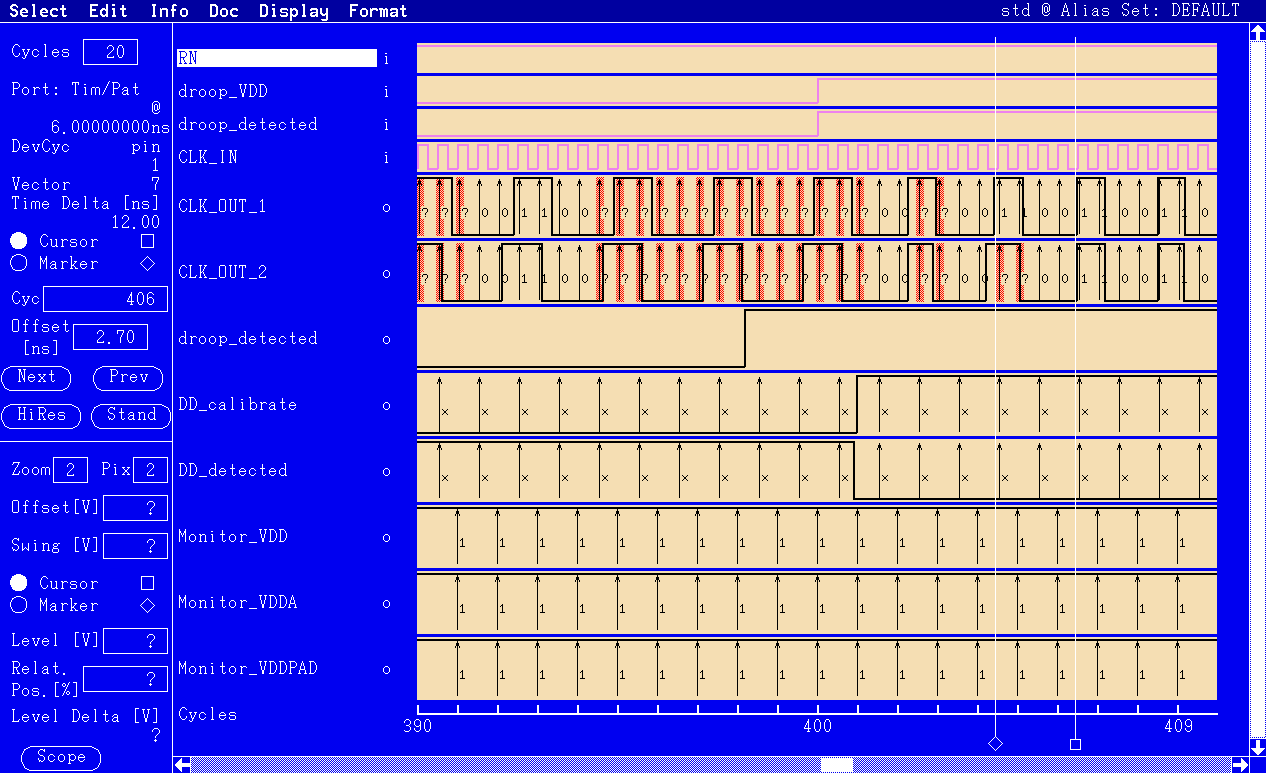}
    \caption{This is a variant of \Cref{fig:noisy_out_bad_frequency} with all digital outputs on display. This figure illustrates the digital operation of our droop detector. We note that the signals \signal{DD\_detected} and \signal{DD\_calibrate show that a droop within a single output clock cycle as seen in \signal{CLK\_OUT1} and \signal{CLK\_OUT2}}} 
    \label{fig:all_digital_droop_detector}
\end{figure*}
\begin{figure*}[p]
    \centering
    \includegraphics[width=.98\linewidth]{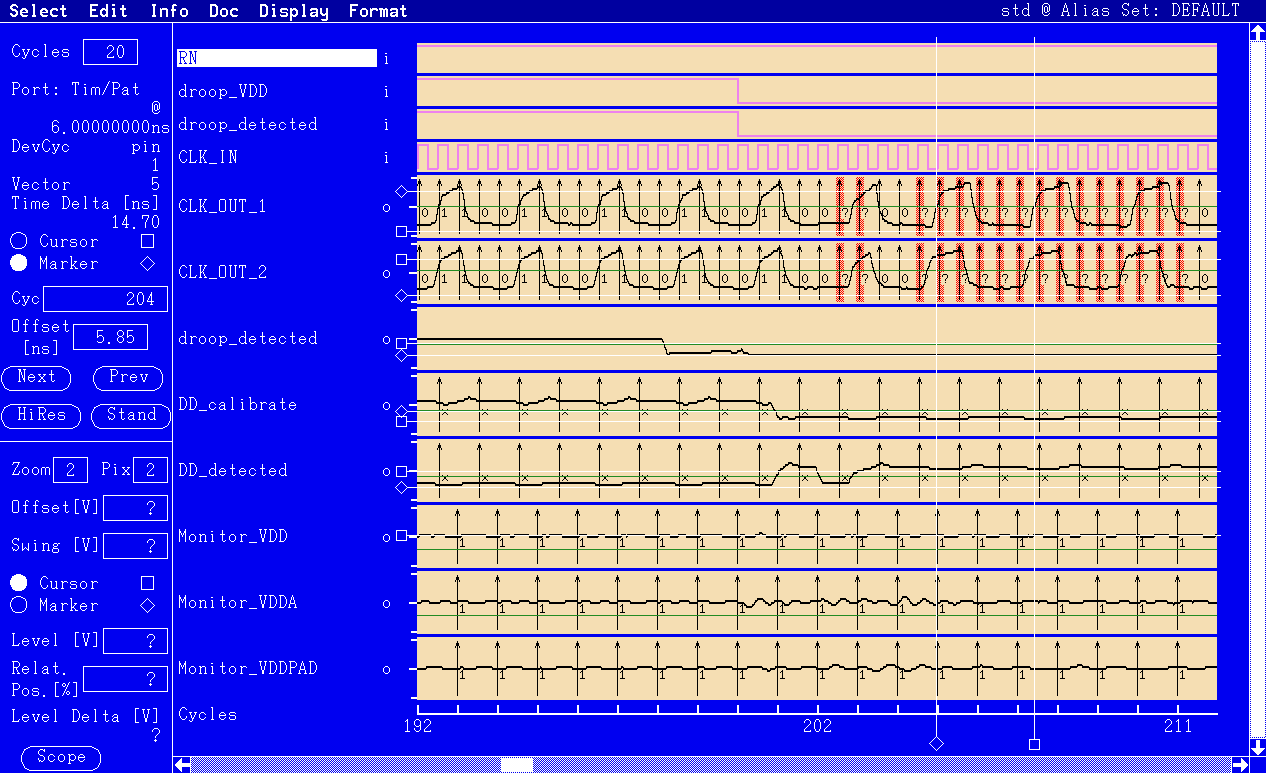}
    \caption{Behavior of FAM modules under under logic \signal{droop\_detected} input. The input \signal{droop\_detected}, for which both analog and digital signals are shown, switches to $0$, indicating a droop. We observe that this triggers an increase in the period of our output clocks, as expected.}
    \label{fig:normal_rise}
\end{figure*}

\begin{figure*}[p]
    \centering
    \includegraphics[width=.98\linewidth]{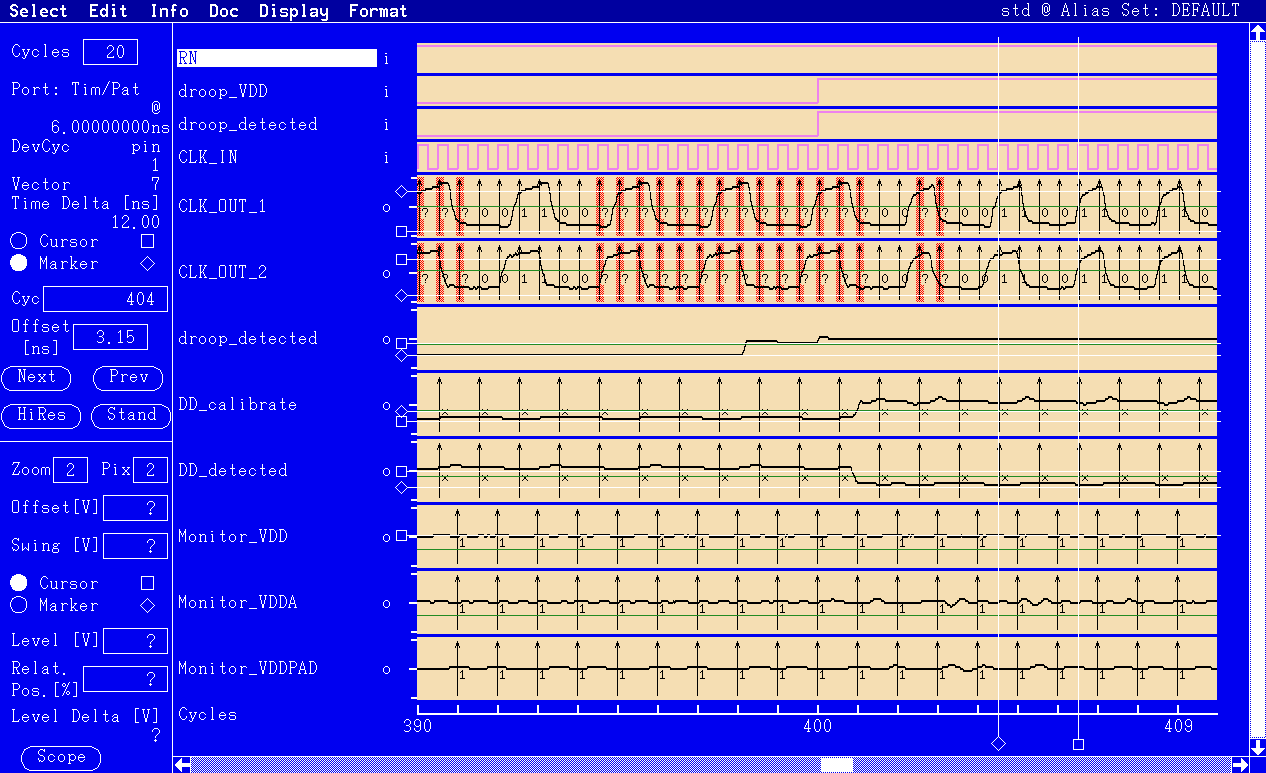}
    \caption{Behavior of FAM modules under under logic \signal{droop\_detected} input. The input \signal{droop\_detected}, for which both analog and digital signals are shown switches to $1$, indicating that a droop ended. We observe that this reverts the period of our output clocks to the nominal one, as expected.}
    \label{fig:normal_fall}
\end{figure*}

\subsection{FAM behavior in standard operation}
\begin{comment}
Recall that we used a common test bench to test the behavior of both \signal{FAM-I} and \signal{FAM-II} with and without metastability. For this sub-section, we focus on the non metastable operation of FAM-I. 
\end{comment}

Let us recall the relevant high-level idea of FAM-I and the test bench. \signal{FAM-I} gets as input a digital signal coming from an analog port \signal{droop\_detected}. This is unlike \signal{FAM-II}, which gets the corresponding input from the droop detector's \signal{DD\_Detected} output. Since we provide a digital input through an analog port, we can control the voltage level of the input at any given time slice. We exploit this feature to construct a unified test bench for \signal{FAM-I} to test it under both normal conditions and non-logic inputs. We can already see an example of this in \Cref{fig:droop-detector-clock-freq-1}, where the analog trace of input \signal{droop\_detected} appears to transition to an intermediate a voltage from the 198th cycle until the 200th cycle of the tester, before dropping all the way to \signal{GND}. By controlling the voltage of the signal in this intermediate level, we can both generate ``clean'' \signal{droop\_detected} inputs and attempt to force the capturing latch into metastability.

For testing the ordinary behavior of \signal{FAM-I}, it suffices to look at the digital value of \signal{droop\_detected}. In \Cref{fig:normal_rise}, we see that the digital value of \signal{droop\_detected} switches to active at cycle 200 of the tester. The effect of the delayed pulses occurs one clock cycle later, as seen in \signal{CLK\_OUT\_1}. We also observe that changing our clock period from the original design does not introduce any glitches. Specifically, the time period of our pulses goes from 12\,ns to 14.7\,ns. Similarly, when the droop detection signal is de-asserted, the clock pulses return to normal operation within a clock cycle as shown in the digital signals output by the tester \Cref{fig:all_digital_droop_detector}. The delay of one clock cycle comes from our mechanisms for safely latching a droop detection signal only after a previous clock signal has passed. This extra clock cycle also proves sufficient for \signal{droop\_VDD} to pass through the droop detector and get sampled. Thus both clocks \signal{CLK\_1} and \signal{CLK\_2} change their frequencies at the same clock cycle.

\subsection{FAM behavior under non-logic \signal{droop\_detected} inputs}
Having tested that our circuit operates well under clean digital inputs, and performed clock frequency measurements under simulations of very long droops, we now subject \signal{FAM-I} to testing under intermediate voltage level \signal{droop\_detected} inputs. A few hurdles present themselves immediately. Our droop detector captures its values in a flip-flop, which would have to not only maintain internal metastability for an extended period of time, but might by chance offer an output voltage that is interpreted as either logic 0 or logic 1 by downstream latches despite internal metastability.

Yet, such a filtering should not be taken as a given and other detector designs might avoid capturing measurements in a flip-flop or latch altogether for the sake of speed. To address this issue, we bypass the droop detector by directly supplying a ``bad'' \signal{droop\_detected} signal. To this end, in our test benches the input signal \signal{droop\_detected} is artificially set to an intermediate voltage, which we call $V_{\mathrm{int}}$, before dropping to $\signal{GND}$ and decisively asserting a droop. We then test our circuit across a range of values for $V_{\mathrm{int}}$ to determine its robustness to bad inputs.

\begin{figure*}[ht]
    \centering
    \includegraphics[width=0.8\textwidth]{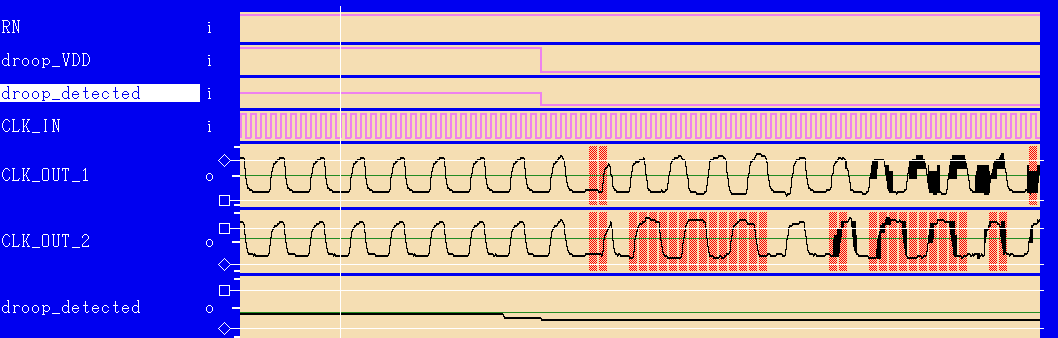}
    \caption{When the active low droop detection input is set to 0.5V, \signal{CLK\_OUT\_1} is observed to slow down indicating that 0.5V is amplified into logic 0 internally. Note that \signal{CLK\_OUT\_2} receives its droop detection input independently from the droop detector}
    \label{fig:step_01_meta_0p5}
\end{figure*}

\begin{figure*}[ht]
    \centering
    \includegraphics[width=0.8\textwidth]{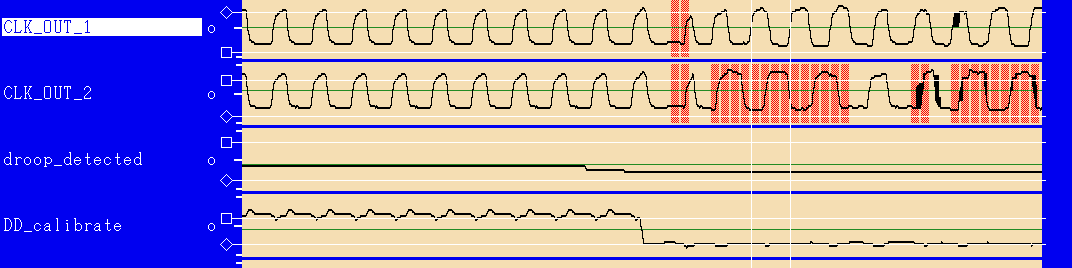}
    \caption{When the active low droop detection input is set to 0.6V, \signal{CLK\_OUT\_1} appears to have the same output frequency as before suggesting that this voltage is translated internally to logic 1}
    \label{fig:step_01_meta_0p6}
\end{figure*}

To find the critical voltage threshold of the latch in question, we first run our test bench by varying $V_{\mathrm{int}}$ at step sizes of $0.1V$. The basic idea is that voltages slightly lower or higher than $V_{\mathrm{int}}$ are quickly resolved as \signal{logic 0} or \signal{logic 1} and the FAM modules output clocks which are unambiguous across the multiple re-runs of the test bench. We illustrate this in \Cref{fig:step_01_meta_0p5} and \Cref{fig:step_01_meta_0p6} for $V_{\mathrm{int}} = 0.5V$ and $V_{\mathrm{int}} = 0.6V$, respectively. First, we compare the analog and digital versions of \signal{droop\_detected} in both figures.


As we approach the voltage where the latch could potentially be driven into deep metastability, the resolution of analog input to digital output after stabilization becomes ambiguous, which translates to noisy clock output. Having narrowed down the voltage range of $V_{\mathrm{int}}$ from $0.5V$ to $0.6V$, we extracted timing diagrams at various settings of $V_{\mathrm{int}}$ at step sizes of $0.01V$ and finally to $0.001V$. At this point, we reached the accuracy limits our testing device and were able to identify $V_{\mathrm{int}}$ to be between $0.574V$ to $0.575V$, likely closer to the former. The output of some of these scans is shown in the figures below 

Next, we fix the voltage $V_{\mathrm{int}}$ at $0.574V$ and run the test benches. At this point, let us again emphasize the following. In our test setup, we cannot distinguish any internal metastability in the latches from noise caused by temperature and voltage variations. In any event, our test results do indicate that our FAM modules are resilient to a wide range of input voltage variations at the droop detected input. In particular, in all our tests the output clocks are glitch-free with clean signal transitions and the test chip reacts to voltage droops within one clock cycle of the occurrence of the droop. 

\subsection{Clock Frequency Measurements under Long Droops}
In our final set of measurements we measure the frequency of the clock as the droop detection input signal's voltage is varied at a granularity of 0.005 V. The choice of the long droop is motivated by two reasons.
\begin{itemize}
    \item On the one hand, our circuit must behave well even in the presence of long droops. 
    \item On the other hand, our tester requires sampling enough clock edges to make a frequency measurement of the output clocks.
\end{itemize}

Some sample results of our measurements are shown in \Cref{fig:droop-detector-clock-freq-1},
\Cref{fig:droop-detector-clock-freq-2}, and \Cref{fig:droop-detector-clock-freq-3}. These represent frequency measurements of the output clock signal during the occurrence of a droop, asserted by the droop detection signal. Because of the fragility of testing for metastability, we instead choose to simulate several potential intermediate voltages that could occur when metastability strikes. Since our tester only accepts digital logic values, we manipulate the voltage of the high impedance logical value \signal{Z}. For each of our test benches, \signal{Z} voltage values, and our dies, we measure the clock frequency of the output clock. We observed in our results, that upto limits of experimental variation, our clocks consistently change frequency when the \signal{Z} is set to a voltage value between 0.5V and 0.6V, indicating the threshold when the droop is asserted. The almost clean frequency transition of the output clock frequency further indicates that our circuit indeed handles a wide range of input voltages on the droop detection input. Our collected data can be found with the supplementary material.

% \appendices
% \section{Placeholder title}
% \TODO{Do we want an appendix? Christoph: To what end?}


\section{Conclusion}\label{sec:conclusion}

In conclusion we implemented and tested the design of \cite{srinivasDesign24} on the IHP 130 nm technology. The circuit responds to droops within 1.2 clock cycles and successfully slows down an 80 MHz clock, despite originally being designed for 100 MHz. For this frequency we were also able to demonstration using a strobe that our circuit produces clean clock signals. An open question that remains is how one might deploy such a chip on a large GALS system, and the synchronisation and response-time trade-offs that arise from the placement of this circuit in a clock tree, closer to the source or the registers.    

\section*{Acknowledgment}
We thank the Max Planck Institute for Informatics for their support during the part of the project conducted when Christoph Lenzen and Shreyas Srinivas were employed there. No AI tools were used in the preparation of this manuscript.

\ifCLASSOPTIONcaptionsoff
  \newpage
\fi

\bibliographystyle{IEEEtran}
\bibliography{references}

\begin{IEEEbiography}[{\includegraphics[width=1in,height=1.25in,clip,keepaspectratio]{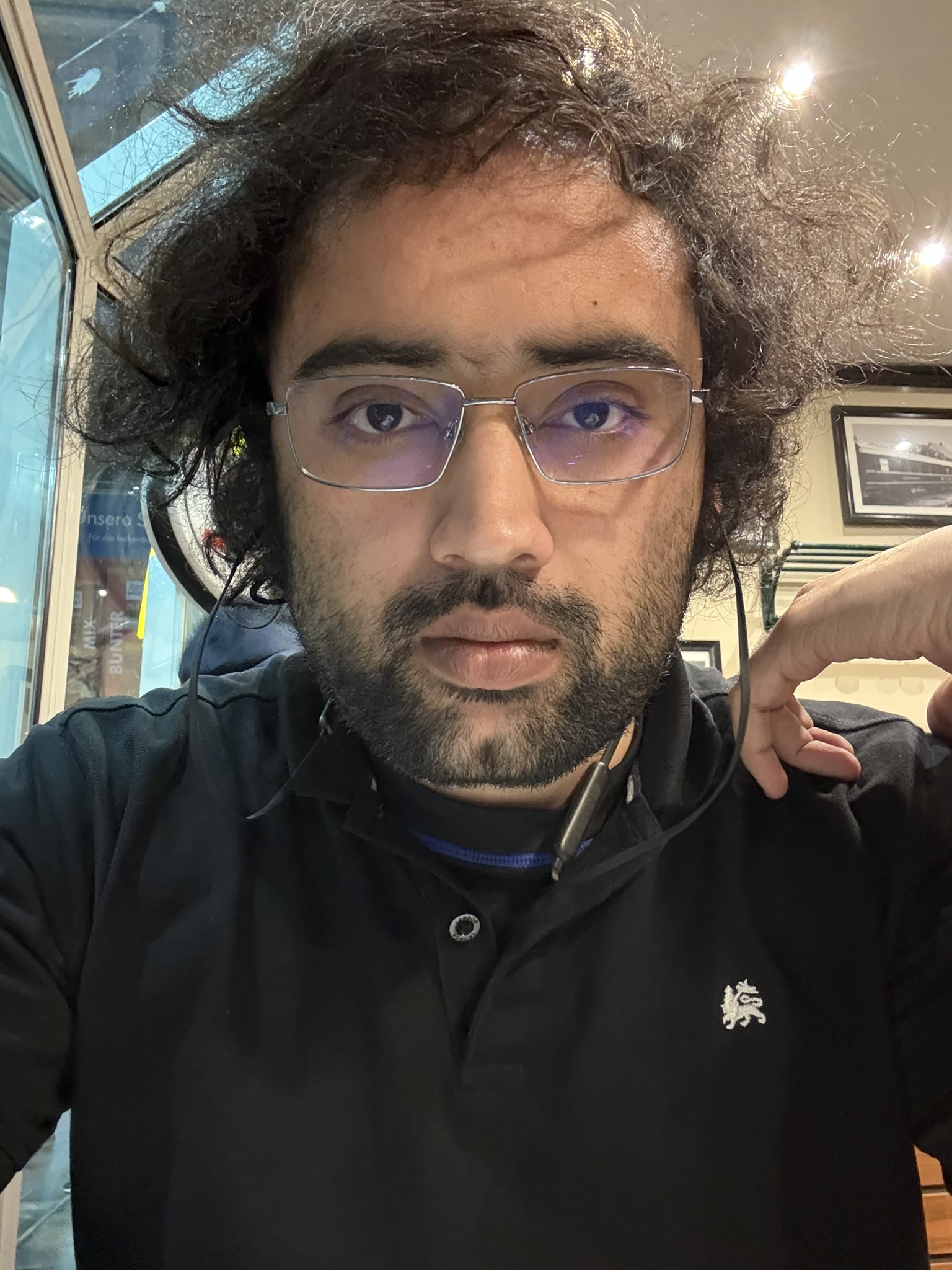}}]{Shreyas Srinivas}
is a PhD student at the CISPA Helmholtz Center for Information Security. Prior to this he received his bachelor's degree from the National Institute of Technology, Trichy, India, in 2016 and his masters degree from the University of Oxford in 2017. His current research interests include the mathematics of clock synchronisation and the use of interactive theorem provers for formal verification. 
\end{IEEEbiography}

\begin{IEEEbiography}[{\includegraphics[width=1in,height=1.25in,clip,keepaspectratio]{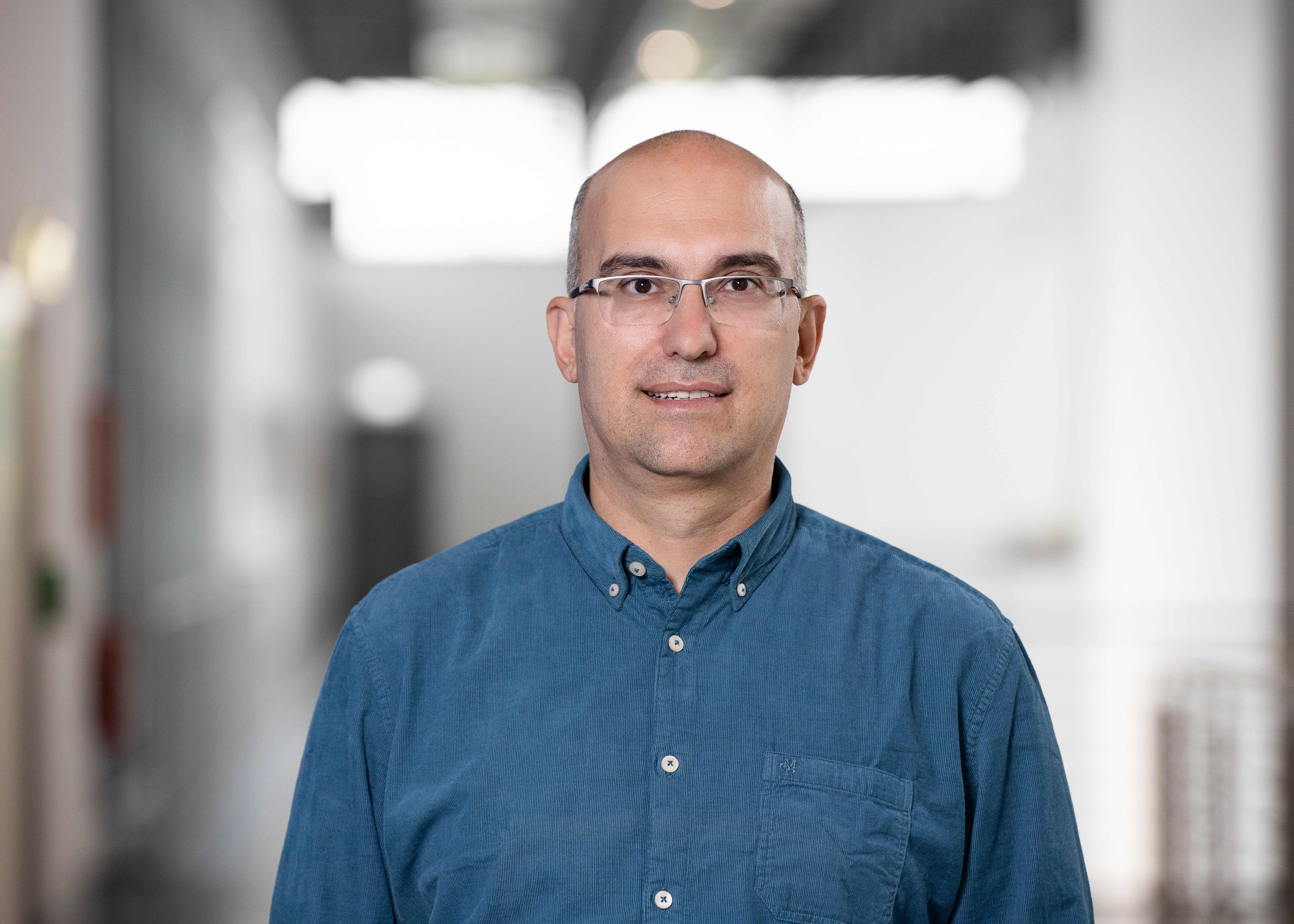}}]{Milos Krstic}
 Prof. Dr. Milos Krstic received the Dr-Ing. degree in electronics from Brandenburg University of Technology, Cottbus, Germany in 2006. Since 2001 he has been with IHP, Frankfurt (Oder), Germany, where he leads the department System Architectures. From 2016 he is also professor for “Design and Test Methodology” at the University of Potsdam. For the last few years, his work was mainly focused on fault tolerant architectures and design methodologies for digital systems integration. Prof. Krstic has been managing many international and national R\&D projects at IHP (GALAXY, EMPHASE, IC-NAO, ENROL, RTU-ASIC, SEPHY, DIFFERENT, VHiSSi, RESCUE, MORAL, BB-KI Chips, etc.). He has published more than 300 journal and conference papers, and registered 12 patents.
\end{IEEEbiography}

\begin{IEEEbiography}[{\includegraphics[width=1in,height=1.25in,clip,keepaspectratio]{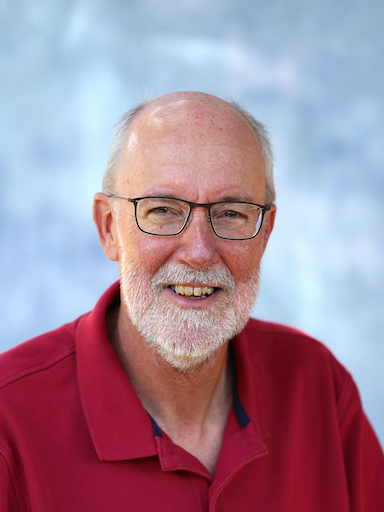}}]{Ian W Jones}
is a circuit researcher, inventor, and developer with expertise
in clock domain crossings, metastability, and asynchronous circuits.
Ian is an IEEE Life Member.
\end{IEEEbiography}

\begin{IEEEbiography}[{\includegraphics[width=1in,height=1.25in,clip,keepaspectratio]{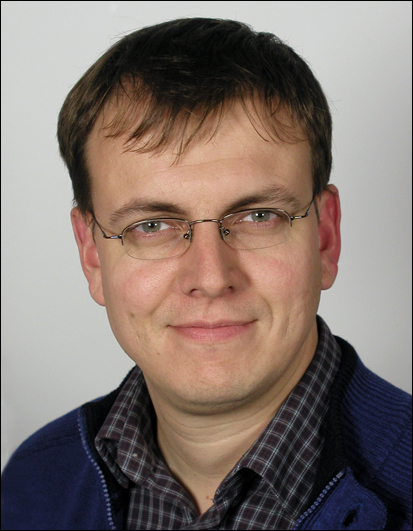}}]{Carsten}
  holds a Diplom degree in Computer Science from Brandenburg University of Cottbus-Senftenberg. Currently he works as a test engineer at the IHP – Leibniz Institute for High Performance Microelectronics where he focus on the development of automated testing methods.
\end{IEEEbiography}

\begin{IEEEbiography}[{\includegraphics[width=1in,height=1.25in,clip,keepaspectratio]{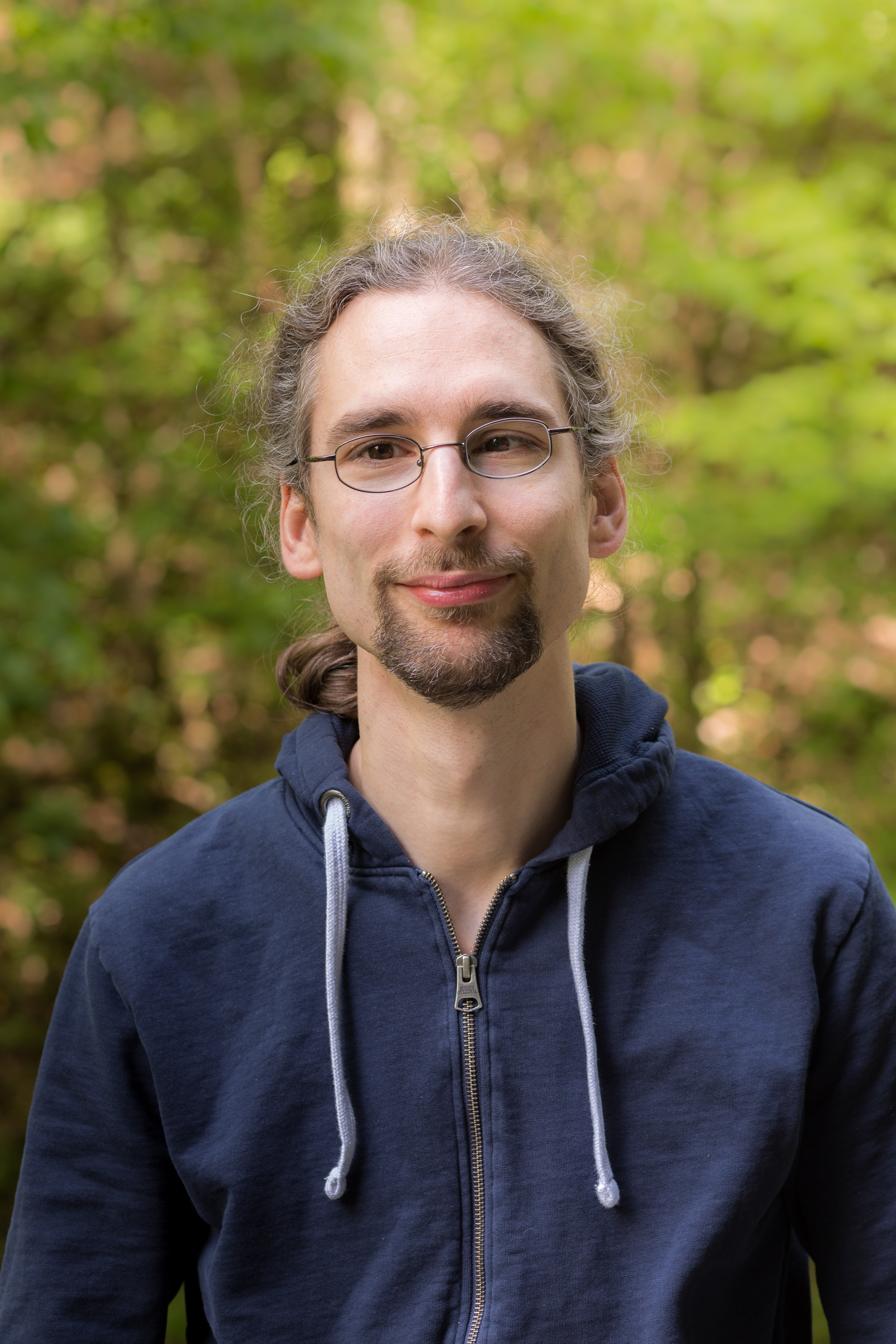}}]{Christoph Lenzen}
Christoph Lenzen received a diploma degree in
mathematics from the University of Bonn in 2007
and a Ph.\ D.\ degree from ETH Zurich in 2011. After postdoc positions at the Hebrew University of
Jerusalem, the Weizmann Institute of Science,
and MIT, he became group leader at MPI for
Informatics in 2014. Since 2021, he is faculty at
CISPA Helmholtz Center for Information Security. He received the best paper award at PODC
2009, the ETH medal for his dissertation, and in
2017 an ERC starting grant.
\end{IEEEbiography}

\vfill


\end{document}